# MANIFESTATION OF STRONG QUADRUPOLE LIGHT-MOLECULE INTERACTION IN THE SEIRA SPECTRA OF SOME SYMMETRICAL MOLECULES

A.M. Polubotko

A.F. Ioffe Physico-Technical Institute Russian Academy of Sciences, Politechnicheskaya 26, 194021 Saint Petersburg RUSSIA, Tel: (812) 292-71-73, Fax: (812) 297-10-17,

E-mail: alex.marina@mail.ioffe.ru

## ABSTRACT

It is demonstrated, that the forbidden lines, which were observed in the SEIRA spectra of diprotonated $BiPyH_2^{2+}$ and ethylene, adsorbed on $Cu$, as well as on ethylene, adsorbed on mordenites can be explained on the base of the Dipole-Quadrupole theory of SEIRA and is the consequence of the existence of the strong quadrupole light-molecule interaction.

Investigation of surface enhanced optical processes is of great importance, since it allows to reveal a number of valuable physical details, which are of great importance for various areas of science. Here we would like to draw attention on the fact, that the forbidden bands, which were observed in the SEIRA spectra of some symmetrical molecules can be explained by a strong quadrupole light-molecule interaction, which exists in these systems [1]. One should remind that the conception of the strong quadrupole light-molecule interaction allows to create the theory of Surface Enhanced Raman scattering (SERS) [1] and the theory of Surface Enhanced Hyper Raman scattering (SEHRS) [2-5]. The most important feature, which proves existence of this interaction, is appearance of forbidden lines in the SER and SEHR spectra of molecules with sufficiently high symmetry. Explanation of their appearance was made in [1-5].

Forbidden bands in SEIRA were observed in several works. In [6] the authors observed forbidden lines, caused by totally symmetric vibrations in the SEIRA spectra of diprotonated $BiPyH_2^{2+}$ adsorbed on copper. Its SEIRA spectra are presented on Figure 1, while the list of the observed lines is contained in Table 1.

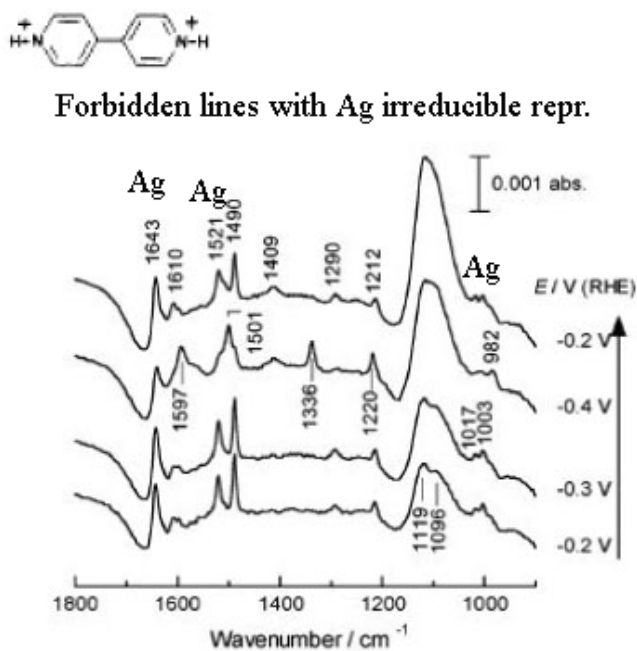

Figure 1. The SEIRA spectra of diprotonated $BiPyH_2^{2+}$ [6].

| SEIRA of $BiPyH_2^{2+}$ The applied potential is (-0.3-0.1 V) | Assignment in the $D_{2h}$ group |
|---|---|
| 1643 (s) | $A_g$ |
| 1610 (w) | $B_{2u}$ |
| 1597 (w) | $B_{2u}$ |
| 1521 (s) | $A_g$ |
| 1490 (s) | $B_{2u}$ |
| 1409 (w) | No assignment |
| 1290 (w) | $B_{2u}$ |
| 1212 (w) | $B_{1u}$ |
| 1017 (w) | $A_g$ |

Table 1. Assignment of the absorption IR and Raman lines of diprotonated $BiPyH_2^{2+}$ adsorbed on copper and in another systems [6].

One can see that several part of the lines, are caused by totally symmetric vibrations transforming after the unit irreducible representation $A_g$ of the symmetry group of this ion - $D_{2h}$, and are forbidden in the usual infrared absorption. The line at 1100 $cm^{-1}$ refers to the spectra of perchlorate, which present in this system [6] and is not included in Table 1. In addition there are the lines, caused by vibrations with $B_{1u}$ and $B_{2u}$ irreducible representations, which are allowed in usual infrared absorption. One should note that assignment to the $B_{1u}, B_{2u}$ and $B_{3u}$ irreducible representations depends on the orientation of the ion with respect to the coordinate system. Regretfully the authors of [6] do not present this information. However in any case existence of the lines with "u" symmetry is in a good agreement with our theory.

Forbidden and allowed lines are strictly defined conception, which is associated with vanishes to zero of matrix elements of quantum transitions and is based on precise theorems of the group theory. However researchers in [6] tried to explain their appearance with charge transfer from molecules to the surface, or back that correspond to disturbance of symmetry due to adsorption. From our point of view this explanation is not satisfactory because of the following reason. The strong charge transfer should cause sufficiently large deflection of the lines from their positions for the free molecules because of the change of the force constants of the molecule, which is bound with the surface. However the deviation of the wavenumbers of the molecules is sufficiently small. For example for $BiPyH_2^{2+}$ in solid (dichloride) or in acidic solution the deviation of the wavenumbers compared with adsorbed $BiPyH_2^{2+}$ is within the interval $\pm 10$ $cm^{-1}$ and is less than 1% from the values of the wavenumbers of these bands, which is nearly the error of their determination. Therefore we can not consider that there is a significant charge transfer between $BiPyH_2^{2+}$ and copper. In addition explanation of appearance of the strong forbidden bands is in contradiction with the idea of continuity of the values of intensities of the forbidden lines, since slight distortion of symmetry can not cause so large growth of the intensities.

In [7], forbidden bands in SEIRA also caused by totally symmetric vibrations were observed for ethylene also adsorbed on copper. Consideration of the work [7] requires some special attention. From our point of view the authors have not given a distinct reason of their appearance. In accordance with results of [7] there are two kinds of lines. The first group of the lines appear at low coverage, approximately equal, or less than 4.7L with sufficiently large intensity. They are the lines at 897, 1276 and 1536 $cm^{-1}$. These lines are absent in the ethylene spectra of the free ethylene molecules. Another group of the lines are those, which appear at the coverage 4.7L, or more. They are the lines at 821, 948, 1220, ,1335, 1436 and 1617 $cm^{-1}$ with smaller intensity and positions, which are close to the ones of the free ethylene (Figure 2). The last ones are 826, 949, 1220, 1344, 1444 and 1630 $cm^{-1}$, in accordance with the data published in [8]. From our point of view the first group of the lines with sufficiently large frequency shifts can be associated with the molecules, adsorbed in the first layer, which interact strongly with the copper surface. Their sufficiently strong intensities and appearance in the experiments with small coverage of the substrate confirm our point of view that these molecules are placed in the first layer. This result is well known in the theory of Surface Enhanced Optical processes and is named as a first layer effect. However, in accordance with the results of [9-11] we can suppose very large distortion of these molecules, from their geometry in a free state. Therefore we may deal with the complex, which symmetry can strongly differ from the symmetry of the free ethylene. Interaction of ethylene with copper can be investigated on the base of the results of adsorption of ethylene on various facets of copper, on Cu (110) in particular [10,11]. In accordance with [9] ethylene can adsorb by two manners, as it is shown on Figure 3. Calculations of the ethylene frequencies, made in [12] for the pointed configurations demonstrate, that there are the most prominent lines in the region of 1532, 1275 and 909 $cm^{-1}$, which originate from the lines at 1630, 1344 $cm^{-1}$ with $A_g$ symmetry and 949 $cm^{-1}$ with $B_{3u}$ symmetry of the free ethylene. Large frequency shifts and appearance of the first two lines in the IRRAS spectra of ethylene on Cu (110) in [10,11] point out that ethylene apparently distorts strongly because of adsorption that results in influence of both the dipole and quadrupole interaction in formation of these lines (the last one can also arise also due to residual roughness, which may present in this system).

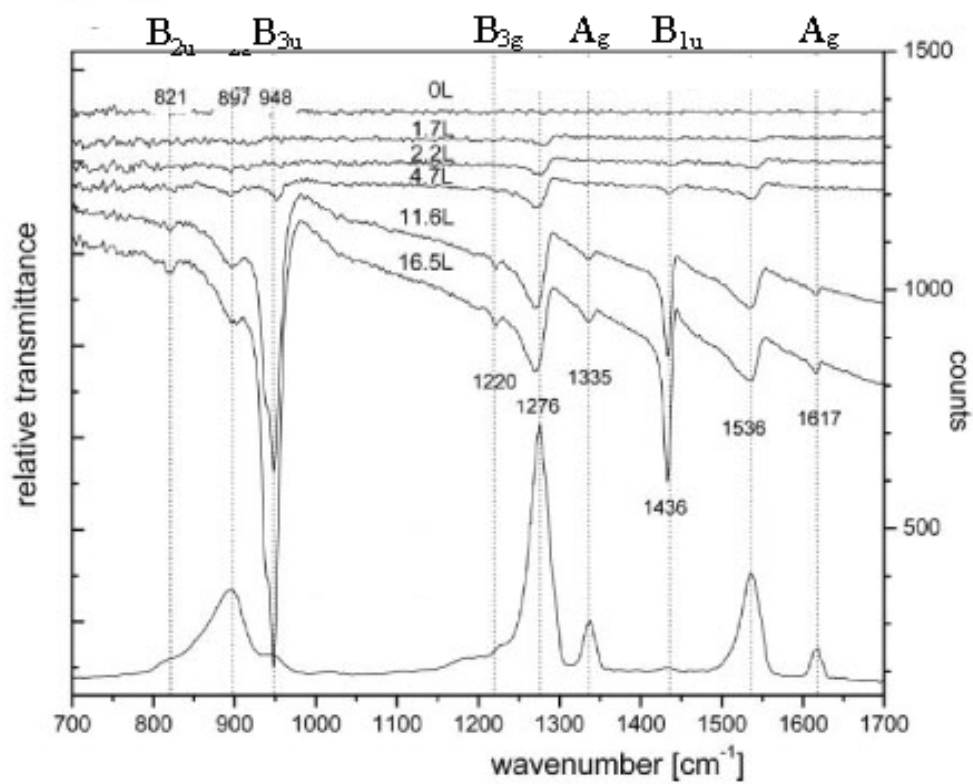

Figure 2. The SEIRA (upper curves) and SER (lower curve) spectra of ethylene, adsorbed on copper [7].

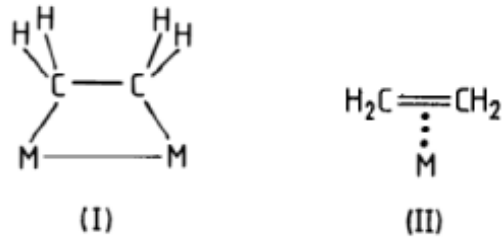

Figure 3. Possible geometry of adsorption of ethylene on copper [9]. Under M we assume the metal atoms.

In addition one should note some strange behavior of the line at 909 $cm^{-1}$ in experiments in [10,11]. In [10] this line is absent in the spectrum of ethylene on Cu(110) with 1L of the coverage, and in [11] it manifests only for the coverage less than 0.2L, or absolutely absent for some another conditions. The behavior of the above pointed lines points out on the strong change of geometry of ethylene, adsorbed on copper. Therefore we can not judge about manifestation of the quadrupole interaction on the base of behavior of the lines at 897, 1276 and 1536 $cm^{-1}$ in [7], since they belong to the strong distorted ethylene molecules adsorbed in the first layer with the strong interaction with copper. However existence of the second group of the lines, especially at 1335 and 1617 $cm^{-1}$ for the coverage larger than 4.7L, which are nearly unshifted with respect to the lines of the free ethylene points out that they arise from the molecules, adsorbed in the second and upper layers since their amplitudes are small and the molecules are situated outside the first layer, where the enhancement must be maximal. Since the charge transfer is absent for these molecules, appearance of the nearly unshifted bands 1335 and 1617 $cm^{-1}$ with the $A_g$ irreducible representation, which are allowed in the usual Raman scattering and in SERS, can be explained by existence of the strong quadrupole light-molecule interaction only, as it will be demonstrated below. One should stress that the quadrupole mechanism of appearance of the forbidden lines does not cause any frequency shifts, while the charge transfer mechanism causes these shifts that is in a good agreement with our ideas.

In addition there are the lines with the wavenumbers 1436, 948 and 821 $cm^{-1}$, which appear at the coverage, which is approximately equal, or larger than 4.7L and refer to the vibrations with $B_{1u}, B_{3u}$ and $B_{2u}$ irreducible representations. (Here the irreducible representations are written out for the molecule with the $x$ axis, which is perpendicular to the molecule plane and the $z$ axis, which is perpendicular to the C=C bond.). These lines are nearly unshifted compared with the lines of the free molecule. They belong to the nearly undistorted molecules and are allowed in SEIRA. One should note appearance of the weak lines at 897 and 1220 $cm^{-1}$. The first one originates from the line 948 $cm^{-1}$ and is strongly shifted due to the strong chemical interaction of the molecules with copper from our point of view. Therefore we have some difficulties to assign any irreducible representation to it. Its appearance can be associated with the absence of a definite symmetry, when both the dipole and quadrupole interactions become allowed. Explanation of appearance of the line 1220 $cm^{-1}$ assigned to the $B_{3g}$ irreducible representations causes some difficulties at present. It is associated with the fact that it is unshifted with respect to the corresponding line of the free ethylene. In addition it must be forbidden in the dipole approximation in SEIRA. Therefore we shall refrain from explanation of its appearance here.

Thus the main peculiarity of the SEIRA spectra is appearance of the forbidden lines at 1335 and 1617 $cm^{-1}$, caused by totally symmetric vibrations transforming after the unit irreducible representation $A_g$. Appearance of the lines, caused by the vibrations with $A_g$ irreducible representation in the SEIRA spectra of diprotonated $BiPyH_2^{2+}$ and ethylene, adsorbed on copper can be explained by existence of the strong quadrupole light-molecule interaction, or by the Dipole-Quadrupole theory of SEIRA which was developed in [13,14]. Below we shall present some simplified aspects of the Dipole Quadrupole theory of SEIRA, which can help in understanding of the reasons of appearance of the forbidden lines. More detailed and precise theory one can find in [13-15].

The precise Hamiltonian of interaction of light with molecules has the form

$$\hat{H}_{e-r} = -\sum_i \frac{ie\hbar}{mc} \overline{A}_i \nabla_i , \qquad (1)$$

where $\overline{A}_i$ is vector potential of the electromagnetic field in the place of the $i$ electron. Summation in (1) is made over all $i$ electrons. All other designations are conventional. The Hamiltonian (1) can be transformed and written in the so called dipole approximation. However one can demonstrate that in our case, for molecules adsorbed near rough metal surfaces it is necessary to take into account the quadrupole terms of the expansion (1). In this case the Hamiltonian has the form

$$\hat{H}_{e-r} = |\overline{E}| \frac{(\overline{e}^* \overline{f}^*)e^{i\omega t} + (\overline{e}\overline{f})e^{-i\omega t}}{2} , \qquad (2)$$

where

$$f_i = d_i + \frac{1}{2E_i} \sum_\beta \frac{\partial E_i}{\partial x_k} Q_{ik} , \qquad (3)$$

$d_i$ and $Q_{ik}$ are the $i$ and $ik$ components of the dipole moments vector and of the quadrupole moments tensor respectively. $\overline{E}$ is the electric field, which affects the molecule. As it was demonstrated in [1] there is an enhancement of the electric field and its derivatives near prominent places of the rough surface with a large curvature. These places of the rough surface can be approximated by models of the surface of a cone form. The behavior of the radial component of the electric field near the tops of these models can be approximated by the following formula

$$E_r \sim \left|\overline{E_{inc}}\right|_{vol} C_0 \left(\frac{l_1}{r}\right)^{\beta}, \qquad (4)$$

where $\left(\overline{E_{inc}}\right)_{vol}$ is the incident field in a free space, $C_0 \sim 1$ is a numerical coefficient, $l_1$ is a characteristic size of the cone (a height for example), $r$ is a distance from the top of the cone, $0 < \beta < 1$ and depends on the dielectric constant and the cone aperture. In accordance with formula (4), the radial component of the electric field is strongly enhanced near the top of the cone. The same can be said about the derivative $\frac{\partial E_r}{\partial r}$. For more realistic form of the roughness, there will be the enhancement of the component of the electric field, which is perpendicular to the metal surface and the enhancement of the field derivatives $\frac{\partial E_i}{\partial x_i}$. In accordance with formula (3), this enhancement causes the enhancement both of the dipole and the quadrupole interactions. Their relation in the electric field near the top of the cone is

$$\frac{\overline{\langle m|Q_{ik}|n\rangle}}{\overline{\langle m|d_i|n\rangle}} \frac{1}{2E_\alpha} \frac{\partial E_i}{\partial x_k} = B_{ik} a \frac{1}{2E_i} \frac{\partial E_i}{\partial x_k} = B_{ii} \frac{a\beta}{r} \qquad (5)$$

for $i = k$.

$$B_{ii} a = \frac{\overline{\langle m|Q_{ii}|n\rangle}}{\overline{\langle m|d_i|n\rangle}} \gg a, \qquad (6)$$

or $B_{ii} \gg 1$. In (6) $B_{ii}a$ is the relation of some mean values of matrix elements of the quadrupole and dipole moments [13,15]. For example its estimation for the molecule KCN gives the value $B_{ii} \sim 50$. For the benzene, pyrazine or pyridine molecules $B_{ii} \sim 200$. For $i \neq k$ $B_{ik} \sim 1$. This property is associated with the fact that $Q_{ii}$ is the values with a constant sign, while $Q_{ik}$ for $i \neq k$ are the values with a changeable sign that strongly decreases $\overline{\langle m|Q_{ik}|n\rangle}$ matrix elements. From (5) one can see that the quadrupole interaction with $Q_{ii}$ values can be more than the dipole one at

$$r < B_{ii}a\beta \qquad (7)$$

from the top of the cone. A simplified form of the cross-section of infrared absorption can be written as

$$(\sigma_{abs})_{surf} = \frac{\pi\omega_{inc}}{hc\varepsilon_0}\frac{\left|\overline{E}_{inc}\right|^2_{surf}}{\left|\overline{E}_{inc}\right|^2_{vol}} \times \sum_p \frac{V_{(s,p)}+1}{2} \times$$

$$\left|\sum_i (P^e[d_i])\overline{e}_{inc,i} + \sum_{i,k}(P^e[Q_{i,k}])\times \frac{1}{2|\overline{E}_{inc}|_{surf}}\frac{\partial E_{inc,i}}{\partial x_k}\bigg|_{surf}\right|^2, \qquad (8)$$

where

$$P^e[f] = \sum_{\substack{l \\ l\neq n}} 2\,Re\,\frac{R_{nl(s,p)}\langle n|f|l\rangle}{(E_n^{(0)} - E_l^{(0)})} \quad . \qquad (9)$$

Here $(\overline{E}_{inc})_{surf}$ is the surface field, which affect the molecule, $(\overline{E}_{inc})_{vol}$ is the incident electromagnetic field in the volume, $V_{(s,p)}$ is the vibrational quantum number of the $(s,p)$ vibration, $s$ numerates the groups of degenerate vibrational states, while $p$ numerates the states inside the group, $f$ designates the dipole and quadrupole moments. Here in (8), for simplicity, the cross-section is written out, when we take into account only the process of absorption, which occurs only via the electron shell. $R_{nl(s,p)}$ are the coefficients, which describe excitation of the $l$ electronic state from the $n$ ground state by the $(s,p)$ vibrational mode, $E_n^{(0)}$ and $E_l^{(0)}$ are the energies of these states. More detailed theory, concerning determination of the cross-section and other characteristics which are present in (8,9), the $R_{nl(s,p)}$ coefficients in particular one can find in [1,13-15]. The $R_{nl(s,p)}$ coefficients determine the symmetry properties of the excited states $l$ [1,13]

$$\Gamma_l \in \Gamma_n \Gamma_{(s,p)} \qquad (10)$$

where the sign $\Gamma$ designates the irreducible representations of the $l$ and $n$ electronic states and of the $(s,p)$ vibrational mode respectively. In accordance with (9) one can see that $P^e[f] \neq 0$, when it obeys selection rule

$$\Gamma_{(s,p)} \in \Gamma_f \qquad (11)$$

Using (11), taking into account that the most enhancement may arise due to the quadrupole interaction with $Q_{xx}, Q_{yy}, Q_{zz}$ moments one can see that the most enhancement experience the lines, caused by vibrations transforming such as the above moments. In molecules with the $D_{2h}$ symmetry group, which are considered above, they are the vibrations with the unit irreducible representation $A_g$. This result explains appearance of strong bands with $A_g$ irreducible representation for diprotonated $BiPyH_2^{2+}$ and for ethylene adsorbed on copper. The fact that the above lines are not strongest for ethylene does not contradict to the above result, since their strength can depend on the roughness degree and the type of the adsorbent metal. In addition the forbidden lines arise from molecules, situated in the second and upper layers, where the quadrupole interaction is depressed. The quadrupole interaction may be sufficiently strong but not stronger than the dipole one in this case.

Another important result is observation of forbidden lines of ethylene, adsorbed on mordenites [16]. The authors observed the lines in the spectral regions of the lines with the wavenumbers $1344 cm^{-1}$ and $1630 cm^{-1}$, which correspond to the totally symmetric vibrations with the unit irreducible representation $A_g$ (Figure 4). Regretfully we do not know the physical characteristics of the substrates in these experiments, which determine the enhancement of the quadrupole interaction. We have in mind the roughness degree and the complex dielectric constants of the mordenites. However it can be demonstrated that the strong quadrupole interaction of light with molecules can arise not only on metals, but on another substances, such as semiconductors and dielectrics with large modulus of a dielectric constant. The radial component of the electric field near the model roughness of the cone type with large dielectric constant, which can correspond to semiconductors or dielectrics can be described by the formula (4), where the $\beta$ value depends on the dielectric constant of the cone. There is a strong change of the electric field near rough surfaces of such substances too, such as near metals. This causes large increase of the electric field derivatives $\frac{\partial E_i}{\partial x_i}$ and hence the enhancement of the quadrupole interaction, which is responsible for the appearance of the forbidden lines, caused by vibrations with $A_g$ irreducible representation. Thus this mechanism can be operative on mordenites and can explain appearance of the forbidden bands in this system also. The fact that the intensity of these lines can be lesser than the intensity of other lines is associated with not very large enhancement of the field derivatives $\frac{\partial E_i}{\partial x_i}$ since their enhancement depends on the roughness degree and the dielectric constant, which are not known for us. One should note, that the forbidden lines experience slight frequency shifts $\sim 15 cm^{-1}$, which are of the order of 1% of the values of the wavenumbers for these lines, for the spectra (a, b and c). For the spectra (d) and (e) the shifts become larger. Moreover their amplitudes become larger too. The above effect may be associated with the charge transfer, when the dipole interaction become more efficient for these lines, because of larger distortion of the symmetry of the adsorbed molecules.

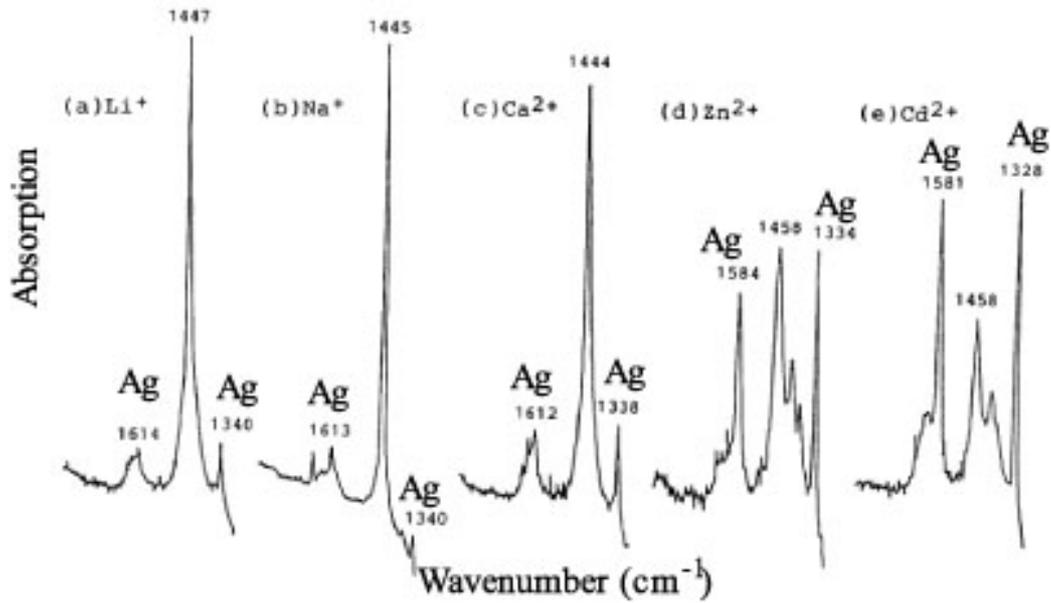

Figure 4. The SEIRA spectra of ethylene, adsorbed on various mordenites [16]. The shifts of the forbidden lines for the (d) and (e) spectra are sufficiently large.

**References:**

[1] A. M. Polubotko, The Dipole-Quadrupole Theory of Surface Enhanced Raman Scattering, 2009 Nova Science Publishers, New York.

[2] A. M. Polubotko, V.P. Smirnov, Journal of Raman Spectroscopy **43**, Issue 3, 380, (2012).

[3] A. M. Polubotko, Optics and Spectroscopy **109,** No. 4, 510, (2010), (English translation of Russian text).

[4] A. M. Polubotko, Journal of Experimental and Theoretical Physics **113**, No. 5, 738, (2011), (English translation of Russian text).

[5] A..M. Polubotko, Chemical Physics Letters **519–520**, 110, (2012).

[6] Yu-Xia Diao, Mei-Juan Han, Li-Jun Wan et al, Langmuir, **22,** 3640, (2006).

[7] Andreas Otto, Walter Akemann and Annemarie Pucci, Israel Journal of Chemistry **46**, 307, (2006).

[8] I. Pockrand, Surface Enhanced Raman Vibrational Studies at Solid/Gas Interfaces, Springer Tracts in Modern Physics, Vol. 104, 1984, Springer-Verlag, Berlin, Heidelberg, NewYork, Tokyo.

[9] N. Sheppard, Ann. Rev. Phys. Chem. **39**, 589, (1988).

[10] C.E. Jenks, B.E. Bent, N. Bernstein and F. Zaera Surface Science Letters **270**, L89,

(1992).

[11] Jun Kubota, Junk N. Kondo, Kazunari Domen, and Chiaki Hirose  J. Phys. Chem., **98**, 7653, (1994).

[12] Koichi Itoh, Tairiku Kiyohara, Hironao Shinohara, Chikaomi Ohe, Yoshiumi Kawamura, and Hiromi Nakai  J. Phys. Chem. B, **106**, 10714, (2002).

[13] A. M. Polubotko, Physics Letters A **173**, 424, (1993).

[14] A..M. Polubotko, Physics Letters A **183**, 403, (1993).

[15] A. M. Polubotko, arXiv:1204.1652 .

[16]  Hidenori Matsuzawa, Hiroshi Yamashita, Masatoki Ito, et al, Chemical Physics **147**, 77, (1990).